\documentclass[preprint,aps,showpacs,preprintnumbers,amsmath,amssymb,superscriptaddress]{revtex4}

\usepackage{graphicx}% Include figure files
\usepackage{dcolumn}% Align table columns on decimal point
\usepackage{bm}% bold math
\begin{document}

%\preprint{APS/123-QED}

\title{Unconditional generation of bright coherent non-Gaussian light from exciton-polariton condensates}
\author{Tim Byrnes}
\affiliation{National Institute of Informatics, 2-1-2 Hitotsubashi, Chiyoda-ku, Tokyo 101-8430, Japan}

\author{Yoshihisa Yamamoto}
\affiliation{National Institute of Informatics, 2-1-2
Hitotsubashi, Chiyoda-ku, Tokyo 101-8430, Japan}
\affiliation{E. L. Ginzton Laboratory, Stanford University, Stanford, CA 94305}

\author{Peter van Loock}
\affiliation{Institute of Physics, Johannes Gutenberg-Universit\"{a}t Mainz, 55099 Mainz, Germany}

\date{\today}% It is always \today, today,

\begin{abstract}
Exciton-polariton condensates are considered as a deterministic source of bright, coherent non-Gaussian light.  Exciton-polariton condensates emit coherent light via the photoluminescence through the microcavity mirrors due to the spontaneous formation of coherence.  Unlike conventional lasers which emit coherent Gaussian light, polaritons possess a natural nonlinearity
due to the interaction of the excitonic component.  This produces light with a negative component to the
Wigner function at steady-state operation when the phase is stabilized via a suitable method such as injection locking.
In contrast to many other proposals for sources
of non-Gaussian light, in our case, the light typically has an average photon number exceeding one and 
emerges as a continuous wave.  Such a source may have uses in continuous-variable quantum information and communication.
\end{abstract}

\pacs{71.36.+c,42.50.Gy,42.50.Ex}% PACS, the Physics and Astronomy
\maketitle

Quantum optics represented by continuous variables \cite{braunstein05,weedbrook12,eisert03,adesso07,paris05} 
has been shown to be a useful approach to
quantum information processing, complementary to the traditional discrete-variable methods using qubits.
While simple, elementary quantum protocols such as teleportation \cite{furusawa98} can be realized using Gaussian
continuous-variable (CV) states and Gaussian operations, any more advanced application such as universal
quantum computation would require some non-Gaussian elements \cite{lloyd99,bartlett02,vanloock11}.  
Moreover, even non-Gaussian states may be insufficient as resources for quantum computation:
trivially, those which are expressible as mixtures of Gaussian states; however, as well those which 
are not representable as mixtures of Gaussian states and yet possess a positive Wigner function \cite{werner95,cerf05}
(for the discrete analogue, see \cite{gross06}). As a consequence, the occurrence of a negative Wigner function
can be seen as a prerequisite for a potential quantum mechanical speed-up \cite{mari12} (for related results
on the discrete case, see \cite{veitch12}). While such negativities are necessary, 
even only small instances of it are sufficient for universality \cite{munronemoto,miranowicz12,mari12}.
Unfortunately, since there are no media with strong enough nonlinearities, current experimental approaches to creating states with a negative Wigner function are highly probabilistic \cite{ourjoum06,neergaard06}, only conditionally preparing such quantum states (those few deterministic schemes using solid-state emitters \cite{eisaman11} would still not be capable of producing bright states). A more desirable way of generating non-Gaussian light is a more ``plug-and-play'' style device, where the device is simply switched on and bright non-Gaussian light emerges deterministically.

In this paper we describe a deterministic method of producing non-Gaussian quantum states of light with a negative Wigner function using exciton-polariton condensates. Exciton-polaritons are bosonic quasi-particles in semiconductor microcavities, corresponding to a coherent superposition of an exciton and a cavity photon \cite{deng10}. Exciton-polaritons have been observed to undergo a dynamical form of Bose-Einstein condensation \cite{kasprzak06,deng02}.  One of the attractive features of this system is that the state of the polaritons inside the
semiconductor can be directly imaged using the leakage of photons through the imperfect microcavity mirrors. Thus the spontaneous buildup of coherence, one of the signals of Bose-Einstein condensation, can be directly measured simply by analyzing the properties of the emerging light.

One of the principle differences between an exciton-polariton condensate and a laser is
the presence, or lack respectively, of strong coupling between the excitons and photons.
Strong coupling leads to polariton-polariton
interactions, giving much richer physics than its non-interacting counterpart. These interaction effects also lead to phenomena such as superfluidity of the polaritons \cite{amo09}. In this paper, we take advantage of the
polariton-polariton interactions to propose a device for producing non-Gaussian light. The key
technological advance that has allowed for this is the availability of high-Q cavities where polariton lifetimes of the order of $\sim 100$ps are now possible \cite{nelsen12}. In addition,
exciton-polaritons may be confined spatially using trapping techniques such as metal deposition
on the surface of the microcavity \cite{kim08}, increasing the self-interaction energy $ U$.  This gives the possibility of energy scales of the self-interaction $ U $ and the cavity decay $ \hbar \gamma_0 $ to be of the same order, leading to the creation of non-Gaussian light in the steady state.
One of the attractive features of this method is that the light is produced completely deterministically and continuously while it emerges from the microcavity.  This is in contrast to most other existing methods which work either conditionally, or with an efficiency that is typically less than 1 for deterministic schemes \cite{eisaman11}. Moreover, the created quantum states have an average photon number much greater than one, thus are much brighter than those from other sources of non-Gaussian light such as single photon emitters.

In the work of Schwendimann {\it et al.}, a master equation for exciton-polariton condensates was
derived, where the ``system'' is considered to be the $ k=0 $ condensate, and the ``reservoir'' are all the modes with $ k \ne 0 $ \cite{schwendimann06,schwendimann08}. The equations are essentially a variation of standard lasing equations, where there are a combination of one and two polariton loss and gain terms, which under suitable conditions gives rise to a condensate population.  In Ref. \cite{schwendimann10} the authors obtain the Fokker-Planck equations for the Wigner functions, and solve the equations numerically and analytically for the approximate case.  The analytical form of the radial dependence of the Wigner function showed no trace of negativity, thus based on this alone it appears that exciton-polaritons cannot be used to generate coherent non-Gaussian light that would be useful for continuous variables.

\begin{figure}
\scalebox{0.40}{\includegraphics{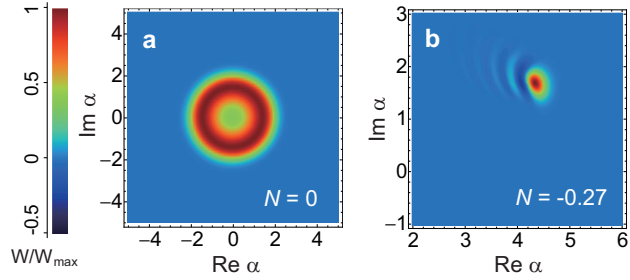}}
\caption{\label{fig1}
The steady-state Wigner function for (a) a laser with no phase fixing $ K = 0 $ (b) an injection locked laser with non-linear interaction $ U = 0.1 $ and $ K = 50 $. Parameters used are $ A = 3 $, $ B = 1 $, $ \gamma = 1 $.  The negativities $ N = \int d^2 \alpha [W(\alpha)-| W(\alpha) |]/2 $ are marked on the plots.  }
\end{figure}

There is, however, an aspect which was not discussed in Ref. \cite{schwendimann10}, which generally applies to all lasing master equations.  For simplicity let us first consider the standard lasing equations, with a hypothetical built-in non-linearity which will be clear that it is analogous to the polariton case. The Scully-Lamb lasing equation \cite{sargent78} with such a non-linearity reads
\begin{align}
\label{lasingmaster}
& \frac{d \rho}{dt} = -\frac{i}{\hbar} [ H_{\mbox{\tiny int}}, \rho ]
-\frac{A}{2} {\cal L} [a^\dagger,\rho] -\frac{\gamma}{2} {\cal L} [a,\rho] \nonumber \\
& + \frac{B}{8} \left[ \rho (a a^\dagger)^2
+ 3 a a^\dagger \rho a a^\dagger  - 4 a^\dagger \rho a a^\dagger a
+ \mbox{H.c.} \right]
\end{align}
where
\begin{align}
H_{\mbox{\tiny int}} & = \frac{U}{2} a^\dagger a^\dagger a a
\end{align}
is the non-linear interaction term, and 
\begin{align}
{\cal L} [O,\rho] \equiv  \rho O^\dagger O + O^\dagger O \rho -  2 O \rho O^\dagger
\end{align}
is a Lindblad loss or gain term, with $ O $ an arbitrary operator. Here
$ A $ is the gain coefficient, $ B $ is the self-saturation coefficient, $ \gamma $ is the cavity loss rate, $ U $ is the non-linear Kerr interaction (self-interaction for condensates). A similar system for a $ \chi^{(2)} $, as opposed to the $ \chi^{(3)} $ nonlinearity as we consider here, was analyzed in Refs. \cite{white96,schack91,sizmann90}.  

The steady-state solution of these equations can be obtained by setting $ \frac{d \rho}{dt} = 0 $. From the
structure of the equation (\ref{lasingmaster}) it can be deduced
that in the steady-state limit only the diagonal components are non-zero \cite{scully97}. This is most easily seen by calculating
the behavior of $ \frac{d P_k}{d t}$, where $ P_k = \sum_n \rho_{n, n+k} $. For diagonal terms $ \frac{d P_0}{dt}=0 $, so that $ P_0 = 1 $ for all times. In contrast, the off-diagonal terms decay away, and in the steady-state limit tend to zero.  Physically,
this can be interpreted as the effect of phase diffusion, where any initial coherence becomes randomized in the long-time limit. Far above threshold, the density matrix of such a state is
\begin{align}
\rho = \frac{1}{2\pi} \int d\theta | r e^{i \theta} \rangle \langle r e^{i \theta} |
= e^{-r^2} \sum_{n=0}^\infty \frac{r^{2n}}{n!} |n \rangle \langle n |
\label{phasediffused}
\end{align}
where $ \alpha =r e^{i \theta} $ and $ | \alpha \rangle $ is a coherent state \cite{rudolph01}.  The state
is diagonal in the density matrix, and the Wigner function has a ``doughnut'' shape with equal amplitude but
a mixture of all phases (see Figure \ref{fig1}a). The point is then that for the diagonal components of the density matrix,
the non-linear term $ H_{\mbox{\tiny int}} $ is completely ineffective, no matter how large $ U $ may be.
This can be seen by looking at the matrix elements
\begin{align}
\langle n | [ H_{\mbox{\tiny int}}, \rho ] | m \rangle = \frac{U }{2} \left[
n(n-1) - m(m-1) \right] \rho_{nm} ,
\end{align}
which disappear for diagonal components. Since diagonal components only couple to other diagonal components in
(\ref{lasingmaster}), the distribution is independent of $ U $.

How then is it possible to make use of the non-linearity?  Clearly we must recover the off-diagonal terms in the density matrix, instead of the completely phase-diffused state (\ref{phasediffused}). This can be achieved by employing a suitable phase fixing method such as injection locking \cite{haus84} or feedback phase stabilization \cite{wiseman93,wiseman94}. The effect of such phase fixing is to take a completely mixed state such as (\ref{phasediffused}), which is immune to the interaction
$ H_{\mbox{\tiny int}} $, and make it approach a pure state depending upon the
strength of the phase fixing. For a pure state it is well known that the interaction
$ H_{\mbox{\tiny int}} $ produces highly non-Gaussian states such as Schrodinger cat states \cite{stobinska08}, and hence we may expect that the interaction will produce non-Gaussian features to the state of the light.

For simplicity, in this paper we consider injection locking to be the phase fixing mechanism.  This corresponds to adding
a coherent term \cite{haus84}
\begin{align}
H_{\mbox{\tiny lock}} & = i \hbar K_0 \left( b a^\dagger e^{-i(\omega - \omega_0)t}  - a b^\dagger e^{i(\omega - \omega_0)t} \right)
\end{align}
to the master equation (\ref{lasingmaster}), so that the first term is $ -\frac{i}{\hbar} [ H_{\mbox{\tiny int}}+H_{\mbox{\tiny lock}}, \rho ] $.  Here $ \omega_0 $ is the frequency of the (slave) laser, and we work in the rotating frame
$ a \equiv a_S e^{-i \omega_0 t} $, where $ a_S $ is the operator in the stationary frame.
$ b $ is the annihilation operator for the injected laser, and $ \omega $ is its frequency which has been factored out.
We assume that bright, on-resonant ($ \omega_0= \omega $) coherent light is used for the injected master laser, so that we may
replace  $ b \rightarrow B e^{-i \psi} $, where $ B > 0 $ is a c-number. Assuming without loss of generality that the injection
locking laser has phase $ \psi = 0 $, and converting (\ref{lasingmaster}) into a Fokker-Planck equation for the Wigner
function, we obtain
\begin{align}
\label{lasingfokker}
\frac{d W}{dt} & = - K \left( \cos \theta - \frac{\sin \theta}{r} \right) \frac{\partial W}{\partial r} \nonumber \\
& + \frac{U}{\hbar} \Big[ (1-r^2) \frac{\partial W}{\partial \theta}  + \frac{1}{16} \Big( \frac{1}{r} \frac{\partial^2 W}{\partial r \partial \theta}  + \frac{\partial^3 W}{\partial r^2 \partial \theta} + \frac{1}{r^2}  \frac{\partial^3 W}{\partial \theta^3} \Big) \Big]  \nonumber \\
&  -\frac{1}{2r}\frac{\partial}{\partial r} \left[ r^2(A-\gamma-Br^2)W \right]  \nonumber \\
& + \frac{(A+\gamma)}{8} \left[ \frac{1}{r} \frac{\partial }{\partial r} \left( r \frac{\partial W}{\partial r} \right)
+ \frac{1}{r^2} \frac{\partial^2 W}{\partial \theta^2} \right]
\end{align}
where $ K = K_0 B $. The steady-state solution of such an equation is shown in Figure \ref{fig1}b.  We see that the phase fixing combined with an interaction $ U $ leads to a steady-state Wigner function with non-Gaussian features, signaled by the presence of a negative region.
This negativity is very stable with respect to initial conditions and remains a feature of the distribution once the steady-state regime is
reached.  Of course, such a laser with a built in non-linearity will in practice have exceedingly small values of $ U $, creating extremely small negativities in the Wigner function.  We shall, however, see that for exciton-polaritons such non-linearities are within experimental reach.

\begin{figure}
\scalebox{0.4}{\includegraphics{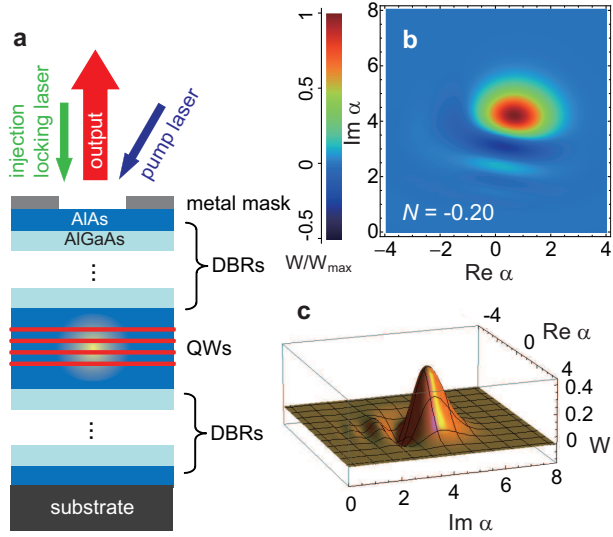}}
\caption{\label{fig2}
(a) The schematic device configuration.  A standard exciton-polariton microcavity with distributed Bragg reflectors (DBRs) is pumped to form a condensate in the quantum wells (QWs). Here we show the example of a AlAs/AlGaAs structure that could be used for the device, although other materials could be used.  An injection locking master laser is injected perpendicularly in order to add a phase fixing term to the condensate.
The results of this paper are concerned with light labeled as ``output'' emerging from the microcavity. (b)(c) Typical steady-state output of the injection locked polariton system.  Parameters used are $ D_2/D_1=10$, $G_1/D_1=1 $, $ G_2=0 $, $ U/D_1 = 10$, $ K/D_1 = 10^3 $.  }
\end{figure}

We first describe our proposed device configuration (Figure \ref{fig2}(a)).  A standard microcavity quantum well system is used, where exciton-polaritons are pumped by an excitation laser or electrical injection \cite{tsintzos09}.  A secondary injection
locking laser is introduced at normal incidence of variable intensity. Metal deposition on the surface of the microcavity allows for a trapping potential which confines the condensate area $ A $. Metal deposition is known to create an effective potential which traps the exciton-polariton condensate \cite{kim08}. The emerging photoluminescence is then used as the output light with non-Gaussian properties. 

To describe this system, our starting point is the theory of Refs. \cite{schwendimann06,schwendimann08}, where a master equation for the polariton condensate is derived. Including the injection locking terms, the density matrix of the condensate evolves in time according to

\begin{align}
\frac{d \rho}{d t} & = -\frac{i}{\hbar} [ H_{\mbox{\tiny int}} +  H_{\mbox{\tiny lock}}, \rho ]
- \Gamma_2 {\cal L} [aa,\rho]
- \Delta_2 {\cal L} [a^\dagger a^\dagger ,\rho] \nonumber \\
&-(\Gamma_1+\gamma_0) {\cal L} [a,\rho]
- \Delta_1 {\cal L} [a^\dagger,\rho].
\label{master}
\end{align}
Here $ a $ is the annihilation operator for a $ k=0 $ polariton, and
$ \omega_0 $ is its frequency (which may include a mean-field energy
shift due to interactions). $ \Gamma_{1,2} $ are one and two polariton
loss terms due to polariton scattering, $ \Delta_{1,2} $ are one and two polariton gain terms due to scattering,
and $ \gamma_0 $ is a loss rate due to leakage through the
microcavity mirrors. From (\ref{master}) it is possible to derive a Fokker-Planck equation for the Wigner function
\cite{schwendimann10}. Using standard techniques \cite{walls08} we obtain
\begin{align}
& \frac{\partial W}{dt} = -2 K \big( \cos \theta - \sin \theta \frac{1}{r} \big) \frac{\partial W}{\partial r}
- \frac{U}{\hbar}(r^2-1) \frac{\partial W}{\partial \theta} \nonumber \\
& +\frac{U}{16 \hbar} \left( \frac{1}{r} \frac{\partial^2 W}{\partial r \partial \theta}
+ \frac{\partial^3 W}{\partial r^2 \partial \theta} + \frac{1}{r^2} \frac{\partial^3 W}{\partial \theta^3} \right) \nonumber \\
& - \left( 2G_2 + 8 G_1 r^2 \right) W  + \left( -G_2 + 2 D_1 - 2 G_1 r^2 \right) r \frac{\partial W}{\partial r} \nonumber \\
& + \left( \frac{D_2}{4} + D_1 r^2 - \frac{G_1}{2} \right) \left[ \frac{1}{r} \frac{\partial W}{\partial r}
+ \frac{\partial^2 W}{\partial r^2} + \frac{1}{r^2} \frac{\partial^2 W}{\partial \theta^2} \right] \nonumber
\end{align}
where we have retained only derivatives up to second order in the Lindblad terms in (\ref{master}), since this gives a very small correction to the overall dynamics. Here $ G_1 = \Delta_1 - \Gamma_1 $, $ G_2 = \Delta_2 - \Gamma_2 - \gamma_0 $, $ D_1 = \Delta_1 + \Gamma_1 $, $ D_2 = \Delta_2 + \Gamma_2 + \gamma_0$ are the net gain and diffusion
coefficients.  The precise expressions for $ G_i $ and $ D_i $ may be calculated using the
expressions in \cite{schwendimann06,schwendimann08,schwendimann10}.  While different materials give rise to
different parameters, the integral expressions given in the above references possess certain properties which
are common to all systems.  As can be deduced from dimensional analysis, the order of the parameters are
approximately
\begin{align}
\Delta_1,\Gamma_1 \sim \frac{V_0^2 A}{4 \pi \hbar E_0 a^2} \hspace{1cm}
\Delta_2,\Gamma_2 \sim \frac{V_0^2 A^2}{8 \pi^3 \hbar E_0 a^4},
\end{align}
where $ V_0 = 6 e^2 a_B / 4 \pi \epsilon A $ is the polariton-polariton interaction energy, $ A $ is the quantization area, $ a $ is a temperature length scale set according to $ E_0 = \hbar^2/2 m_{\mbox{\tiny exc}} a^2 = k_B T $, where $ T $ is the temperature, and $ a_B $ is the exciton Bohr radius. We can observe that the order $ O(\Delta_1/\Delta_2) \sim O(\Gamma_1/\Gamma_2) \sim 2 \pi^2 a^2/A $
is typically a small parameter such that $ \Delta_1,\Gamma_1 \ll \Delta_2,\Gamma_2 $. However, $ \Delta_1,\Gamma_1 $ cannot be neglected as
without these condensation is unstable. Meanwhile the interaction can be estimated to be \cite{byrnes10}
\begin{align}
U \sim \frac{30 e^2 a_B |X|^4}{\pi^3 \epsilon A},
\end{align}
where $ X $ is the excitonic Hopfield coefficient.  From the expressions in Ref. \cite{schwendimann06} it can be deduced that $ G_1<0$ and $ D_{1,2} >0 $, but
$ G_2 $ can change sign which can be taken to be the criterion for condensation assuming $ \Delta_1 \ll G_2 $ \cite{schwendimann06}.

\begin{figure}
\scalebox{0.4}{\includegraphics{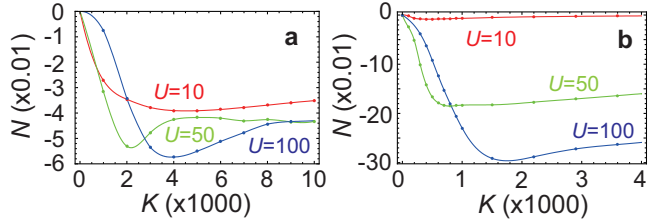}}
\caption{\label{fig3}
Dependence of the total negativity $ N $ with injection locking strength $ K $ for various interactions $ U $ (in units of $ D_1$) for the polariton BEC. Steady-state values are taken for (a) estimated experimental parameters $D_2/D_1=100$,  (b) very high Q cavity/small trap case $ D_2/D_1=10$.  Other parameters are $G_1/D_1=1 $, $ G_2=0 $.}
\end{figure}

In Figure \ref{fig2}(b) and (c) we plot the typical Wigner distribution for the steady-state operation of the injection locked polariton BEC.  The distribution typically consists of a
bright peak that is phase fixed due to the injection locking.  Adjacent to the peak, there is a region of negativity, approximately in the direction of origin.  The negativity appears as soon as the injection locking is switched on, and at first grows with both injection locking strength $ K $ and interaction strength $ U $. For all parameters $ K>0, U>0 $ we see stable negativities of the Wigner function, quantified by $ N = \int d^2 \alpha [W(\alpha)-| W(\alpha) | ]/2 $.  Figure \ref{fig3}a shows the behavior of the amount of steady-state negativity with $ K $. At large enough $ K $, there is a saturation of the negativity, beyond which the total negativity no longer increases, and in fact decreases slowly beyond this point. Although the overall negativity decreases for very large $ K$, the Wigner distribution tends to become more sharply defined for large $ K $. For smaller values of $ K $, the negative region tends to be shallower, but distributed over a larger area. In Figure \ref{fig3}b we plot the negativities
for $ D_2/D_1 = 10 $, which corresponds to very small ($r=0.16\mu$m) traps or very high-Q cavities giving very long polariton lifetimes ($ 1/\gamma_0 = 1$ns).  We see that the reduced phase diffusion improves the negativity of the Wigner function.  We generally also observe that close to threshold gives the largest negativities.

In conclusion, we have proposed a device for deterministically producing bright coherent non-Gaussian light from an exciton-polariton condensate. It was shown that steady-state generation of such light is possible using currently accessible experimental parameters when a suitable method of phase fixing is imposed on the polariton condensate.  Although we have considered injection locking in this paper, other methods such as feedback control of the phase should accomplish the same task. The phase fixing overcomes the phase diffusion which tends to dephase the condensate and leads to a completely mixed state; with phase fixing, non-Gaussian states are created exploiting the natural nonlinearity of the polaritons.  The completely deterministic nature of the production of these non-Gaussian states, as well as their brightness, is potentially useful in the context of continuous-variable quantum information processing, for which at least some resource states must possess a negative Wigner function. On-demand sources will save quantum memories and help scalability, while mean photon numbers
above unity allow for making use of larger Hilbert spaces encoded into the infinite-dimensional space of an optical mode.

T.B. thanks Howard Wiseman, Terry Rudolph and Hui Deng for discussions. This work is supported by the Special Coordination Funds for Promoting Science and Technology, Navy/SPAWAR Grant N66001-09-1-2024, MEXT, and the JSPS through its FIRST program.

%%%%%%%%%%%%%%%%%%%%%%%%%%%%%%%%

\end{document}